\documentclass{ws-jai}
\usepackage[flushleft]{threeparttable}
\usepackage{color}
\begin{document}

\catchline{}{}{}{}{} % Publisher's Area please ignore

\markboth{A. Williamson}{A cosmic ray detection Mode for the Murchison Widefield Array}

\title{An Ultra-High Time Resolution Cosmic Ray Detection Mode for the Murchison Widefield Array}

\author{A. Williamson$^{1}$, C.W. James$^{1}$, S.J. Tingay$^{1}$, S.J.~McSweeney$^{1}$, S.M.~Ord$^{1}$}

\address{
$^{1}$International Centre for Radio Astronomy Research, Curtin University, Bentley, WA 6102, Australia, alexander.williamson1@postgrad.curtin.edu.au\\
}

\maketitle

\corres{$^{1}$Corresponding author.}

\begin{history}
\received{(to be inserted by publisher)};
\revised{(to be inserted by publisher)};
\accepted{(to be inserted by publisher)};
\end{history}

\begin{abstract}
The radio-wavelength detection of extensive air showers (EAS) initiated by cosmic-ray interactions in the Earth's atmosphere is a promising technique for investigating the origin of these particles and the physics of their interactions. The Low Frequency Array (LOFAR) and the Owens Valley Long Wavelength Array (OVRO-LWA) have both demonstrated that the dense cores of low frequency radio telescope arrays yield detailed information on the radiation ground pattern, which can be used to reconstruct key EAS properties and infer the primary cosmic-ray composition.
%can be adapted to observe the sub-mircosecond bursts of coherent radiation from these events.
Here, we demonstrate a new observation mode of the Murchison Widefield Array (MWA), tailored to the observation of the sub-microsecond coherent bursts of radiation produced by EAS. We first show how an aggregate 30.72\,MHz bandwidth (3072\,$\times$ 10\,kHz frequency channels) recorded at 0.1\,ms resolution with the MWA's voltage capture system (VCS) can be synthesised back to the full bandwidth Nyquist resolution of 16.3\,ns. This process, which involves `inverting' two sets of polyphase filterbanks, retains 90.5\% of the signal-to-noise of a cosmic ray signal. We then demonstrate the timing and positional accuracy of this mode by resolving the location of a calibrator pulse to within 5\,m. Finally, preliminary observations show that the rate of nanosecond radio-frequency interference (RFI) events is 0.1~Hz, much lower than that found at the sites of other radio telescopes that study cosmic rays. We conclude that the identification of cosmic rays at the MWA, and hence with the low-frequency component of the Square Kilometre Array, is feasible with minimal loss of efficiency due to RFI.
\end{abstract}


\section{Introduction}

The digital era of radio astronomy has led to a revival of the radio-detection of cosmic rays \citep{HuegeReview,FrankReview}. Extensive air showers (EAS) initiated by cosmic-ray interactions in the atmosphere produce coherent bursts of low frequency ($\sim100$\,MHz) radio-wave radiation over a localised ground region of a few hundred metres diameter. The dense cores of modern low-frequency astronomical telecopes are well-suited for observing this ground pattern from cosmic rays in the $10^{16}$--$10^{18}$\,eV energy range, and provide complementary information to dedicated cosmic ray facilities such as the Pierre Auger Observatory \citep{AERA_Energy_Estimation} and Tunka-Rex \citep{Tunka-Rex}. LOFAR, the Low Frequency Array \citep{2013A&A...556A...2V}, has demonstrated that studying the ground pattern
yields precise information on the primary cosmic rays, and can be used to study their spectrum and, hence, their origin \cite{LOFAR_detecting_cosmic_rays, LOFAR_xmax, LOFAR_composition}. The Owens Valley Long Wavelength Array (OVRO-LWA) has also recently developed a cosmic-ray detection mode \citep{OVRO-LWA}, and cosmic-ray observations with the low-frequency component of the Square Kilometre Array, SKA1-Low, have been proposed \citep{SKA_cosmic_rays}. However, cosmic ray signals appear similar to a bandwidth-limited impulse, so that data analysis ideally requires access to baseband data sampled at the Nyquist rate (direct measurements of the voltage induced at the antennas by the radio waves). This makes it necessary for telescopes aiming to detect cosmic rays to develop a dedicated, high-time-resolution observation mode.

The Murchison Widefield Array (MWA) is a radio interferometer located at the Murchison Radio-astronomy Observatory (MRO) in the Mid West of Western Australia \cite{MWA_phaseI}. Following an upgrade, the MWA consists of 256 tiles, of which 128 can be used concurrently, spread across an area greater than 20 km$^2$ \citep{2018PASA...35...33W} . Each tile consists of 16 dual polarisation antennas arranged in a grid, which are analog beamformed prior to digitisation. The MWA operates between 80 and 300 MHz, with a bandwidth of 30.72 MHz. Its location in a pristine radio-quiet environment, layout, and frequency range make it well-placed to study the radio pulses from EAS initiated by cosmic-ray interactions in the Earth's atmosphere \cite{MWA_phaseII_science}.  To this end, a prototype particle detector has already been installed at the MRO to be used as a trigger \citep{SKAPA_particle_detector}, similar to LOFAR's LORA detector \citep{LORA}.

The MWA can return 30.72\,MHz of processed bandwidth, comparable to the effective bandwidth of LOFAR. This makes cosmic-ray detection with the MWA an attractive experimental prospect, especially since it has been estimated that the 100-200\,MHz range provides the greatest sensitivity to cosmic-ray signals \citep{Pevatron}.
Key to implementing  cosmic ray detection with the MWA is the reconstruction of a time series equivalent to a Nyquist sampled 30.72 MHz bandwidth, starting from the 3072$\times$10 kHz channels produced by the default MWA analog/digital signal chain.

This paper describes a new processing pipeline for the MWA, focused on the detection of cosmic ray signals, which performs this reconstruction process. In Section~2, we briefly describe the properties of the polyphase filterbanks (PFBs) that perform the coarse and fine channelisation steps in the MWA signal chain, and the previously published process for inverting the fine PFB. We then describe our methods for inverting the coarse PFB, and present the expected response of this process to cosmic ray signals. In particular, we focus on the challenges faced when only 24 of 256 channels are available for inversion, and our solution. Section~3 describes an on-site experimental validation of this mode using a calibration pulse, which demonstrates that we have obtained our desired inverse-bandwidth resolution. Section~4 then provides a characterisation of the radio-frequency interference (RFI) environment at nanosecond timescales at the MRO. We present our conclusions in Section 5.

\section{Polyphase Filterbank (PFB) and inversion implementation}

\begin{figure}
    \centering
    \includegraphics[width=0.7\textwidth]{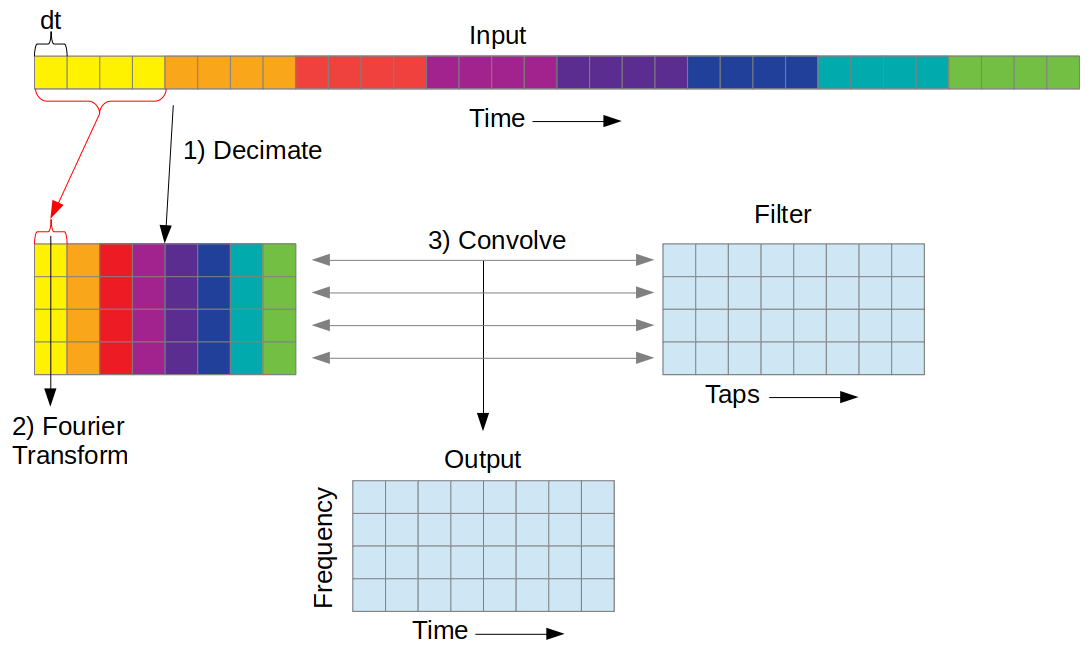}
    \caption{Schematic of a PFB process, using the frequency domain representation of the filters. 1) An initial time-domain data stream at rate $dt$ is decimated into blocks of length equal to the number of desired channels, $N_{\rm chan}$. 2) A discrete Fourier transform (DFT) is then applied along the channel axis. 3) The data are convolved with a forward filter to produce a channelised form of the input data with the time resolution reduced by a factor equal to half the number of channels formed. The inverse process is similar, with the exception that the convolution and (inverse) DFT stages are swapped.}
    \label{fig:pfb_sketch}
\end{figure}

\begin{figure}
    \centering
    \includegraphics[width=0.9 \textwidth]{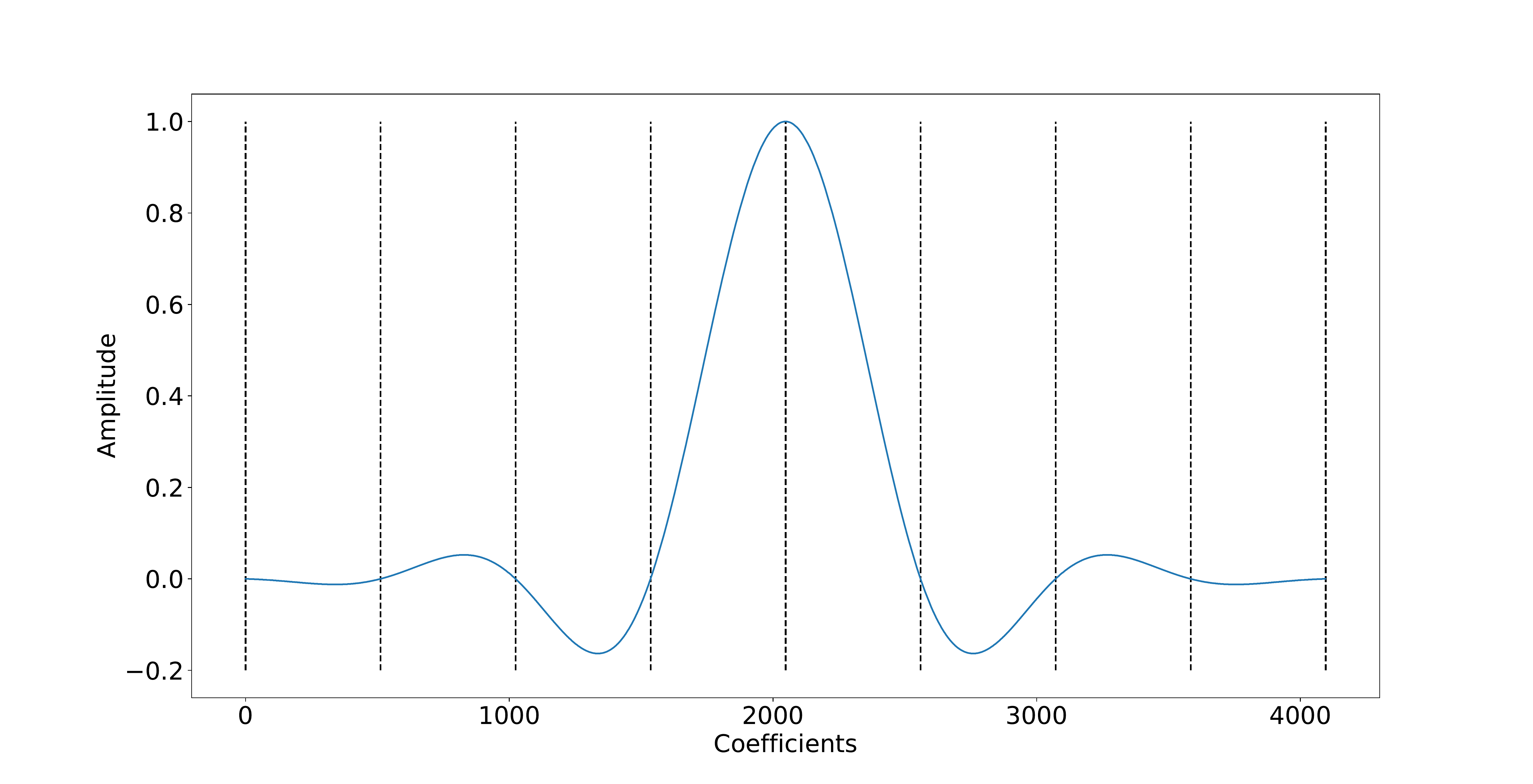}
    \caption{The FIR filter used for both the coarse PFB and its inversion. It is a sinc function convolved with a Hanning window and is split into eight taps of 512 samples each.  The amplitude is unitless and normalised to unity.}
    \label{fig:filter}
\end{figure}

PFBs are commonly used in radio astronomy to convert time series voltages measured from an antenna into their conjugate frequency space in a fast and efficient manner, while minimising spectral leakage between the resulting frequency channels.
The process can be generalised to a decimator that first separates the signal into multiple phases that each undergo a finite impulse response (FIR) filter and a Fourier transform (FT), and are then summed across the phases \citep{MSDP_Crochiere}. The result is a time series for each frequency channel.
Since the FT and FIR filter operations commute, the FIR can be represented in the frequency domain and applied after the FT. This representation of a PFB is shown in Figure~\ref{fig:pfb_sketch};  the time-domain representation of the coarse channel FIR filter is shown in Figure~\ref{fig:filter}.

MWA tile-beamformed voltages for each X and Y polarisation are sampled at 655.36 Msamples s$^{-1}$ (time resolution of 1.53 ns), and channelised using a PFB into 256x1.28 MHz `coarse' channels \citep{2015ExA....39...73P}. This gives a time resolution of 781.3 ns, with 5 real bits and 5 imaginary bits. This first PFB stage uses eight taps of 512 samples each. Of these coarse channels, 24 are returned for central processing, and are passed through a fine channeliser, producing a total of 3072x10kHz channels with a time resolution of 100 $\mathrm{\mu}$s \citep{2015PASA...32....6O}.  This second PFB stage uses 12 taps of 128 samples each.

The MWA voltage capture system (VCS) allows these fine channels to be recorded directly to disk, which is the observation mode commonly used for pulsar observations \citep{VCS}. One second of VCS observation produces 3072 frequency channels with 10$^4$ time samples, for each of 128 MWA tiles for 2 polarisations, resulting in 7.3 GB of data. This means that a one hour observation produces a non-trivial 25.7 TB of data.

In order to search for and reconstruct cosmic ray signals with inverse bandwidth resolution, this large quantity of data must be returned to inverse-bandwidth time resolution by inverting both fine and coarse PFBs, and then searched for cosmic-ray signals.

The inversion process to reconstruct coarse channels from fine filterbank data is described in \citet{MWATiedArrayProcessingIII}, and has already been validated on pulsar observations \citep{2019ApJ...882..133K}. The inverse PFB process is qualitatively similar to the operation of the PFB itself: the data are Fourier transformed back to the time domain, convolved with a FIR filter, and unfolded. Since the PFBs present in the MWA are critically sampled, it is not generally possible to perfectly reconstruct the input time series from the fine channel voltages, due to significant leakage of power between fine channels near the coarse channel edges. Thus we do not investigate the inversion method described by \citet{Morrison2020} and used for fast radio bursts detected by the Australian Square Kilometre Array Pathfinder \citep{Cho2020}, which is based on an oversampled PFB. Therefore, our aim is to choose an inverse filter to reconstruct the input data as accurately as possible.

\citet{MWATiedArrayProcessingIII} investigate several possible inverse filters, finding that the reconstruction accuracy of the coarse channels improves with the number of taps used for the inverse filter. In our application, however, reconstruction of the time-domain data will be limited by the small number of coarse channels available for the inversion process. We therefore use the `mirror filter' described by \citet{MWATiedArrayProcessingIII} for the inversion of both fine and coarse channel data. That is, the forward and inverse filters are identical.

\subsection{Effect of finite channels}

The accuracy of reconstructing a cosmic ray signal with MWA data is primarily limited by the availability of only 24 of 256 coarse channels for further processing. The simplest treatment is to set all unavailable coarse channels to zero prior to the coarse PFB inversion. However, this will cause significant oversampling in the output and will cause the processing to be dominated by calculations on empty elements.
This effect can be negated by choosing the 24 coarse channels to fit into the same Nyquist zone. Fourier transforming only the coarse channels in that zone --- including complex conjugation if it is an odd zone --- will produce a time-domain signal with minimal oversampling. For dedicated observations, the 24 channels can be chosen to lie exactly within a 30.72\,MHz Nyquist zone, resulting in time-domain data at 61.44\,Msamples\,s$^{-1}$. In general, a larger band must be chosen, e.g.\ a Nyquist zone 32 coarse channels wide corresponds to a 1/16 downsampling of the original time-domain data to a rate of 81.2\,Msamples\,s$^{-1}$, with 75\% band occupancy.
Only the FIR filter coefficients corresponding to the used channels are applied prior to the Fourier transform.

We reproduce real-valued time-domain data using a complex-to-real inverse FFT from {\sc FFTW} \citep{FFTW}. The data are then aligned into the time series and written to disk. The output consists of a file for each second, tile, and polarisation.

\subsection{Performance} \label{section:demo}

\begin{figure}
    \centering
    \includegraphics[width=\textwidth]{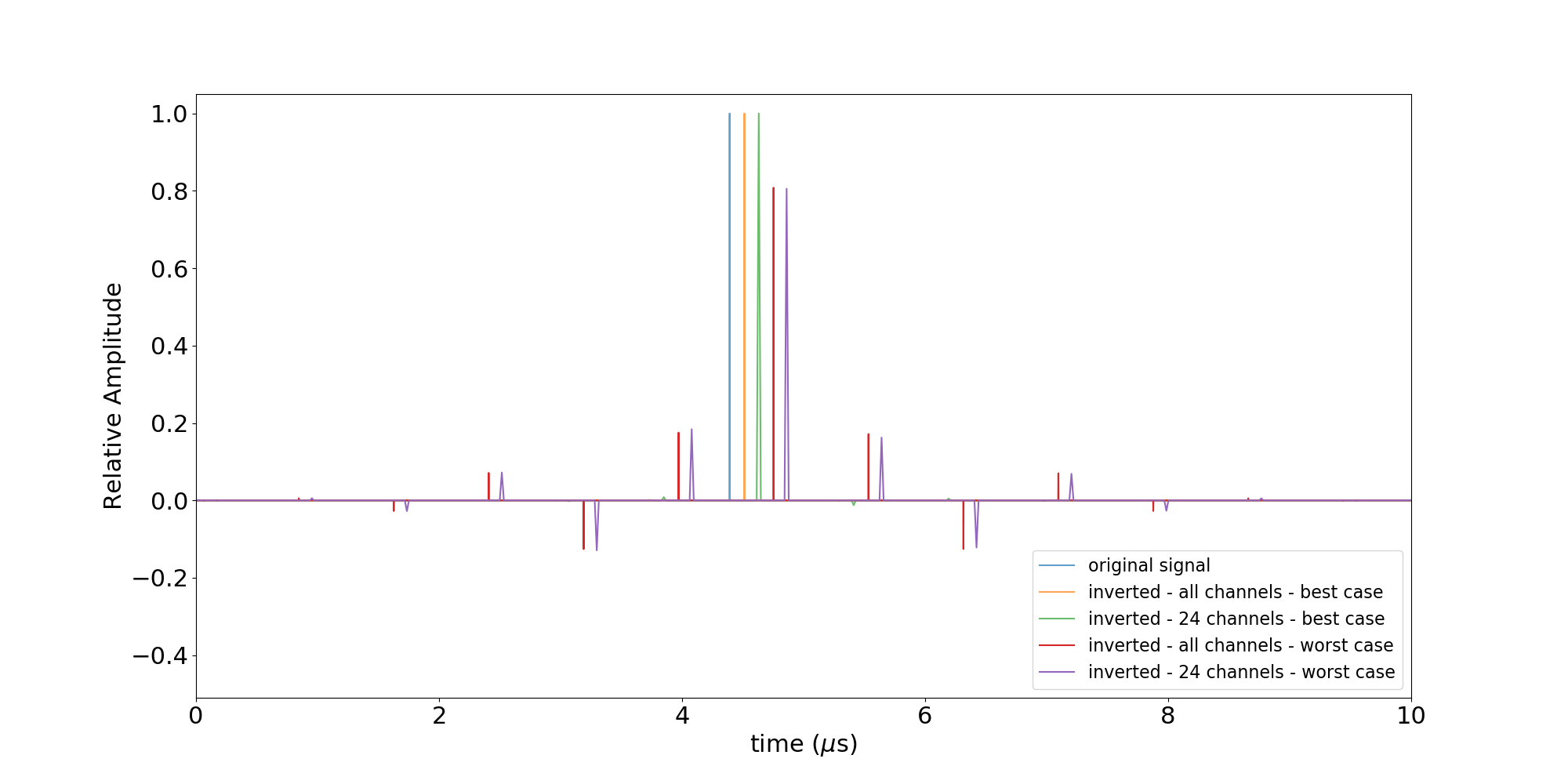}
    \caption{Comparison between the input and reconstructed impulses after being passed through the PFB and inverse-PFB process using all 256 channels and a set of 24 channels. Two locations in the input data stream relative to the PFB tap coefficients are used, corresponding to best-case and worst-case reconstructions. The plots have been offset slightly in order to better display the output signals.}
    \label{fig:comparison}
\end{figure}

Figure \ref{fig:comparison} presents the effect of our PFB inversion on a test impulse passed through software that emulates the MWA coarse and fine filterbanks. The signal is then reconstructed as described above using all 256 coarse channels (i.e.\ no downsampling), and compared to the input. The same signal is then band-limited to 24 coarse channels (96--119), and inverted as described in the previous section.

\begin{figure}
    \centering
    \includegraphics[width=0.5\textwidth]{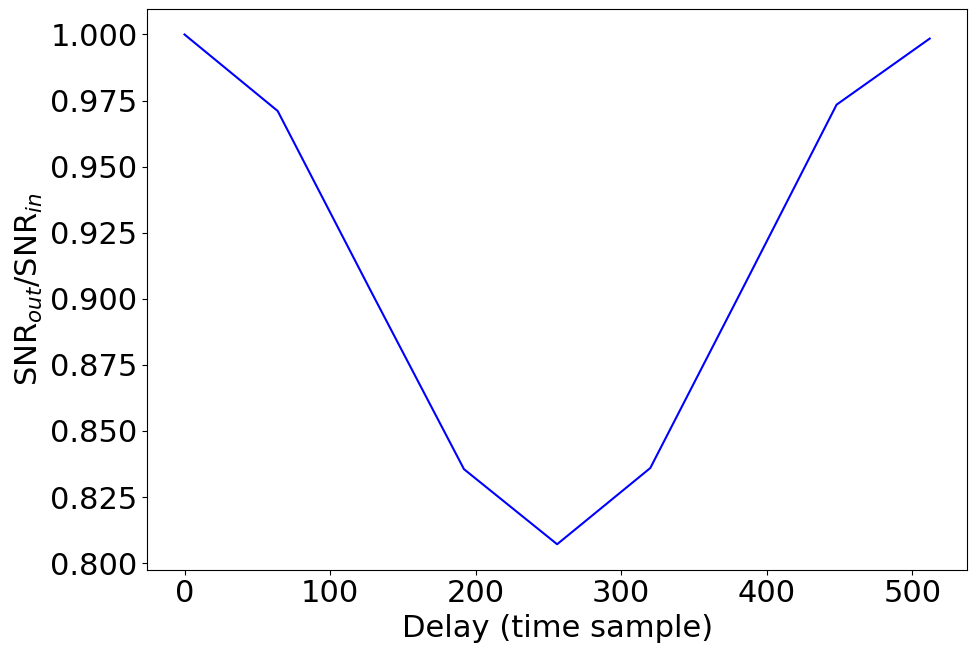}
    \caption{Reconstructed S/N of an impulse when using 256 coarse channels as a function of its location in the original data stream relative to the tap length of the coarse PFB.}
    \label{fig:average_performance}
\end{figure}

The performance of this process varies with the location of the initial impulse within the 512-sample tap length. When the signal is located near the edge of the tap, reconstruction using 256 channels is almost perfect, since PFB coefficients are dominated by the central branch. In the middle of the tap however, the performance is worst, since the PFB coefficients are spread out over the branches. The best and worst cases shown in Figure~\ref{fig:comparison}, respectively, correspond to impulses at the edge and centre of the tap. The performance as a function of tap position is shown in Figure~\ref{fig:average_performance}.

From Figure~\ref{fig:comparison}, the 256 channel inversion in the best case produces a perfect impulse. In general however, the peak of the signal is reduced in amplitude, with secondary peaks showing up as artefacts separated by the tap length of $0.78125\,\mu$s. This results in a reduction in pulse height (which is the signal-to-noise ratio, S/N, when using a peak detection method) of up to 19\%. The reduction is similar when limited to a 24 channel band. The average of the 256-channel performance when varying the location of the impulse in the initial data stream (as per Figure~\ref{fig:average_performance}) is 90.5\%. Since electric field peak height is almost proportional to primary cosmic ray energy \citep{AugerRadioEnergy2016}, the cosmic ray spectrum above the knee of $dN_{\rm CR}/dE_{\rm CR} \sim E_{\rm CR}^{-3.1}$ \citep{KascadeCRSpectrum2012} implies a reduction of 18\% in event rate compared to an idealised experiment.

\section{Demonstration with a pulse calibrator}

In order to test the effectiveness of the inverse PFB, an experiment was conducted at the MRO during scheduled maintenance. A hand-held barbeque gas lighter, which generates a spark via a piezoelectric, was used to produce radio pulses that were both strong enough to be detected by an MWA tile, and with sufficient structure on nanosecond scales to test the cosmic-ray mode.

The gas lighter was triggered a number of times at multiple locations near the core of the MWA during a ten-minute period. Dual polarisation data from coarse channels 96--119 (122.88--153.6\,MHz) were collected from all 128 tiles in the MWA's compact configuration using the MWA VCS and processed through the inverse PFB. This produced a data set with which to validate the high-time-resolution mode.

\begin{figure}
    \centering
    \includegraphics[width=0.8 \textwidth]{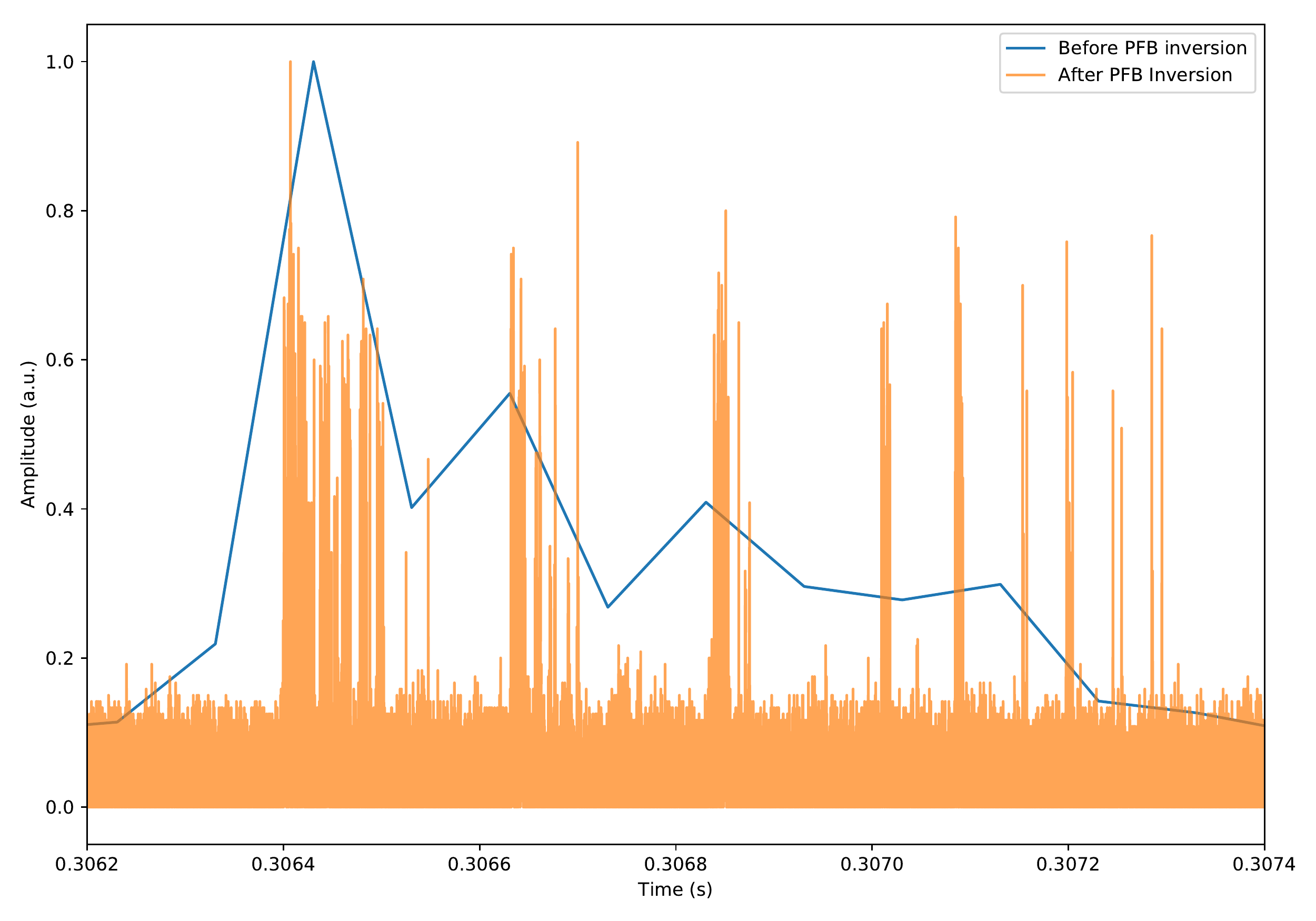}
    \caption{A comparison between the data obtained directly from the VCS (blue) and after the data has been processed to 16.3\,ns time resolution by the inverse PFB software (orange). The high-time-resolution data show significantly more detail, revealing many narrow pulses that are averaged out in the channelised data.}
    \label{fig:clicker_comparison}
\end{figure}

An example of a measured gas-lighter signal is shown in Figure \ref{fig:clicker_comparison}. In order to identify the signal by eye, the unprocessed VCS (fine channel) data were integrated over all tiles and polarisations in 1\,ms intervals. This showed a typical total signal duration of a few ms.
Following the inverse PFB, the signal is prominent in a single tile's time series, and it is possible to discern multiple sub-pulses with microsecond structure.

An automated pulse detection algorithm was implemented to identify gas lighter signals that are coincident in time between multiple tiles. Initially, a simple 5.5\,$\sigma$ threshold search is used on each tile/polarisation independently. The time of arrival of each signal detected at each tile is compared to the signals from all other tiles, and signals that are coincident in time are grouped. The coincidence condition requires that the signals exceed the 5.5\,$\sigma$ threshold within the light travel time between tiles, with an additional 17\,$\mu$s margin (corresponding to the light travel time across the array), after adjusting for system delays. Groups with four or more coincident tiles are noted for further analysis.

%\subsection{Timing and localisation}
%col localisation
% \subsection{Localisation}
%getting delays 

A key difference between cosmic-ray and gas-lighter signals is the total signal duration. The time at which a signal with an intrinsically long duration --- such as that from the gas lighter --- passes above threshold can vary according to signal shape and the influence of random noise. Any signals noted in the coincidence test are therefore correlated to extract a more robust time delay between the signals, resulting in a time resolution comparable to the sampling time. In the case of cosmic-ray signals, which have a duration of a few samples at most, any above-threshold signals will always have an accurate time-of-arrival.

%this can be localised analytically using 4 antennas (2d solution)
It is possible to calculate the origin of a nearfield signal via an analytic approach provided that at least 4 different tiles detect the signal --- three tiles yield two degenerate solutions on the horizontal plane. Since individual tiles can yield large timing errors --- the structure of the gas-lighter signal is such that an incorrect peak in the correlation function is sometimes used to calculate time delays --- candidate source positions for all combinations of four tiles are calculated.
%Due to the expected scatter in the delays measured at the tiles. It is useful to generate a source position solution for all combinations of the tiles that detected the signal. This can give quite wild solutions and it is important to make sure that the tiles that have detected the signal in both polarisations are not included twice.
The median location of all of these calculated points then gives a stable solution to the source location. 

\begin{figure}
    \centering
    \includegraphics[width=0.5\textwidth]{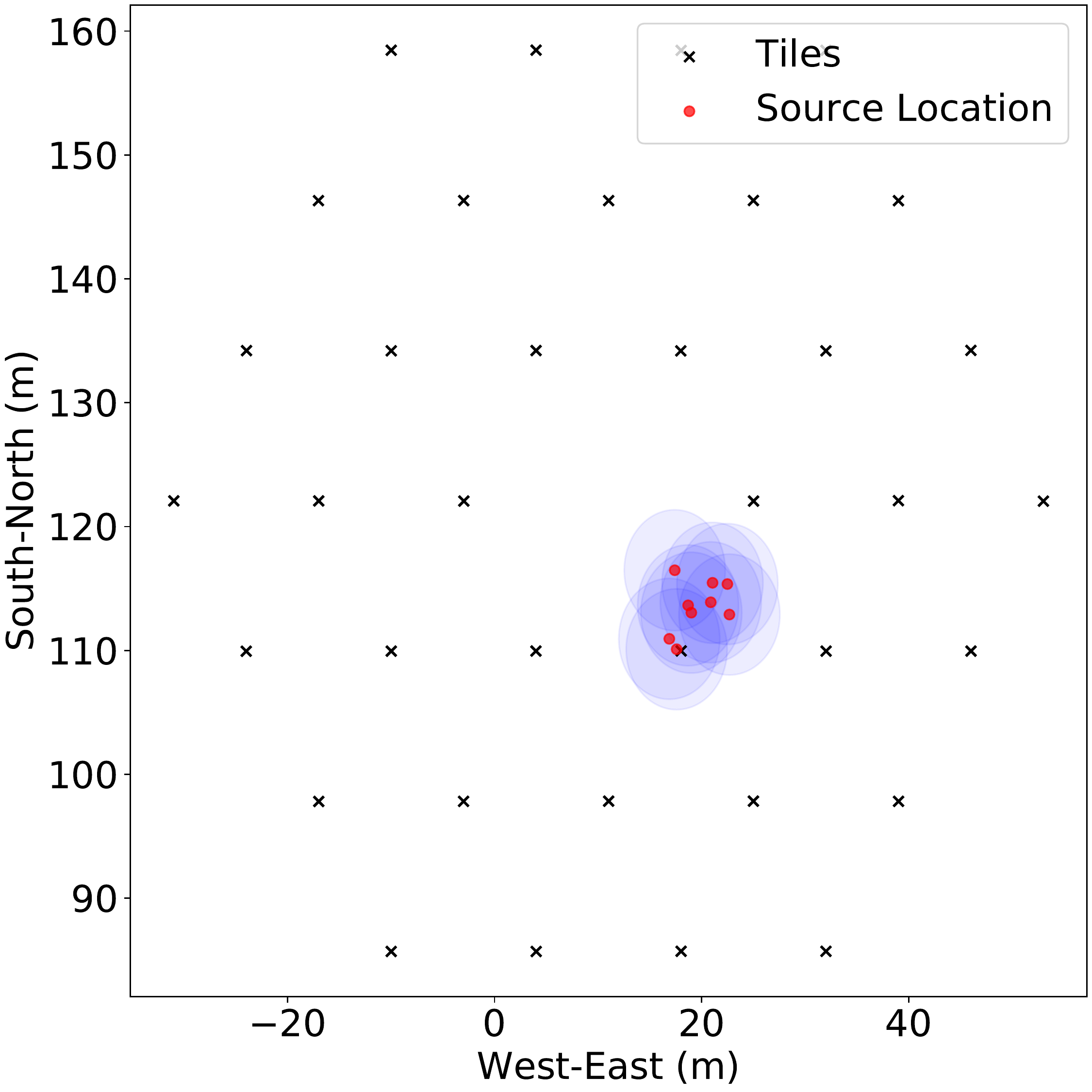}
    \caption{The calculated source locations of a BBQ lighter via the measured radio pulses from MWA tiles. Shown are tile locations for part of the compact configuration of the MWA (black), the calculated source locations (red) and the measurement uncertainty from the sample rate from each tile (blue).}
    \label{fig:localisation_scatter}
\end{figure}
%both methods yield similar results with a scatter of ~5m
%causes for the scatter

Figure \ref{fig:localisation_scatter} shows the final source locations for multiple pulses that occurred within the same second, i.e.\ with the same source location. The scatter of the location solutions is about 5m, consistent with the expected uncertainty due to the resolution of the time series data, and the known location of the gas lighter at the time.

\section{High Time Resolution RFI at the MRO}

One of the major challenges encountered in identifying cosmic ray signals is eliminating impulsive RFI events. The nanosecond-scale RFI environment at the MRO to which this observation mode is sensitive is largely unknown, since it would be averaged out during normal MWA operating modes and would only contribute to a slightly higher noise level. We therefore present first results for low-frequency RFI detected at the MRO, during a one-hour observation. %This project provides an insight into the RFI sampled at nanosecond timescales.

%Our RFI detection method first identified all samples with magnitude greater than six times the root-mean-square (RMS) voltage in each tile/polarisation.

\begin{figure}
    \centering
    \includegraphics[width=0.8\textwidth]{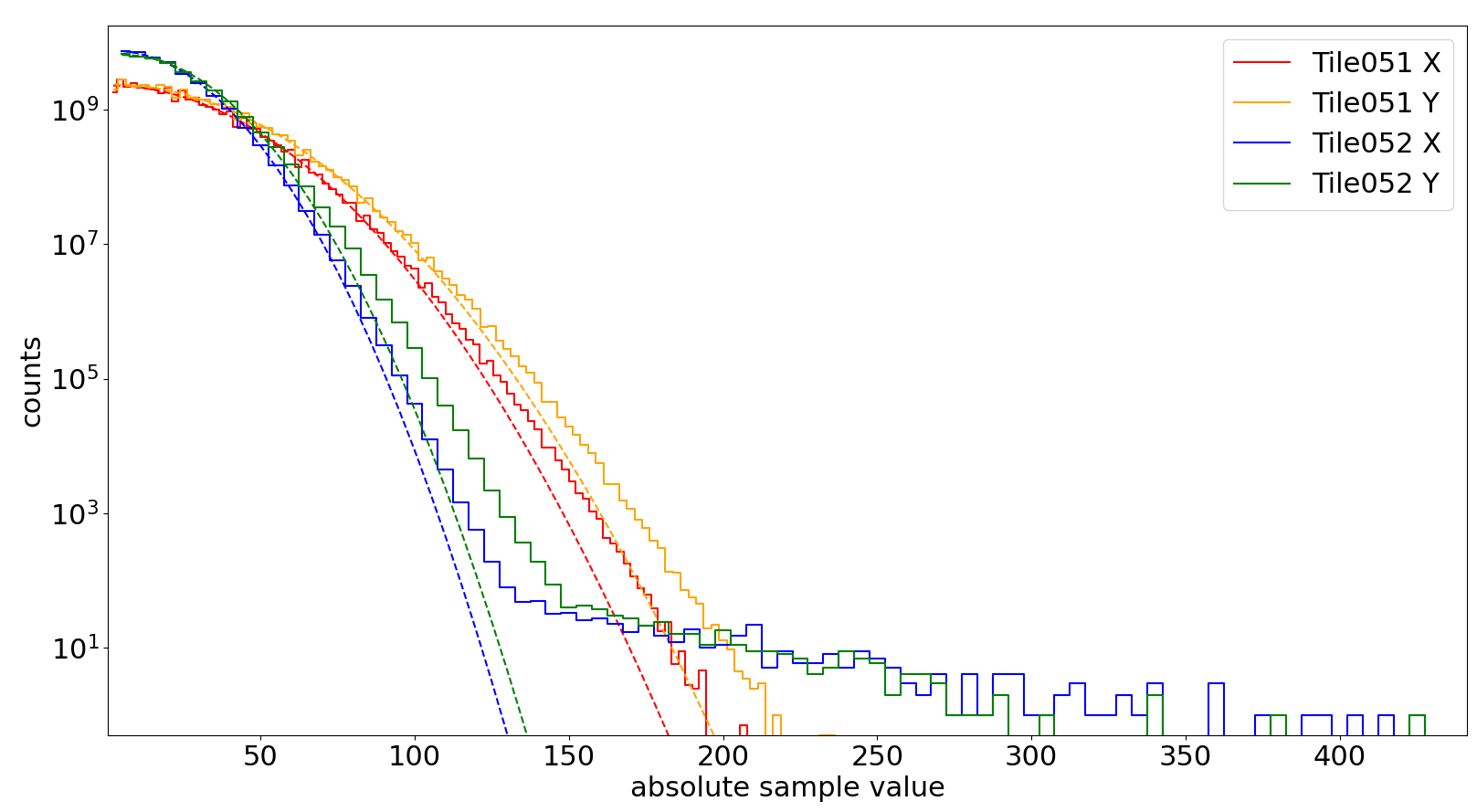}
    \caption{A example of the sample distribution of two core tiles for both polarisation (E-W and N-S) measurements for 442 seconds. The dashed lines represent Gaussian fits to the individual datasets. This figure demonstrates the amount of RFI (i.e. deviations from the fit) detected by the array.}
    \label{fig:gaussianness}
\end{figure}

We first establish the fidelity of our data by analysing the reconstructed voltage amplitudes. Figure \ref{fig:gaussianness} shows two examples of the sample amplitude distributions at each tile. These have been split in X and Y polarisation (orientated East-West and North-South, respectively) and have been fit with a Gaussian curve. The distribution on most tiles/polarisations (e.g.\ Tile 51) appears broadly Gaussian over the full time range analysed. However, all tiles begin to show an excess of samples above about 60--90 digital sample units. RFI events induce a long tail in the distribution, as is evident from the data for Tile 52. %and it is expected that the samples above these values consist of a small amount of RFI.

\begin{figure}
    \centering
    \includegraphics[width=0.6\textwidth]{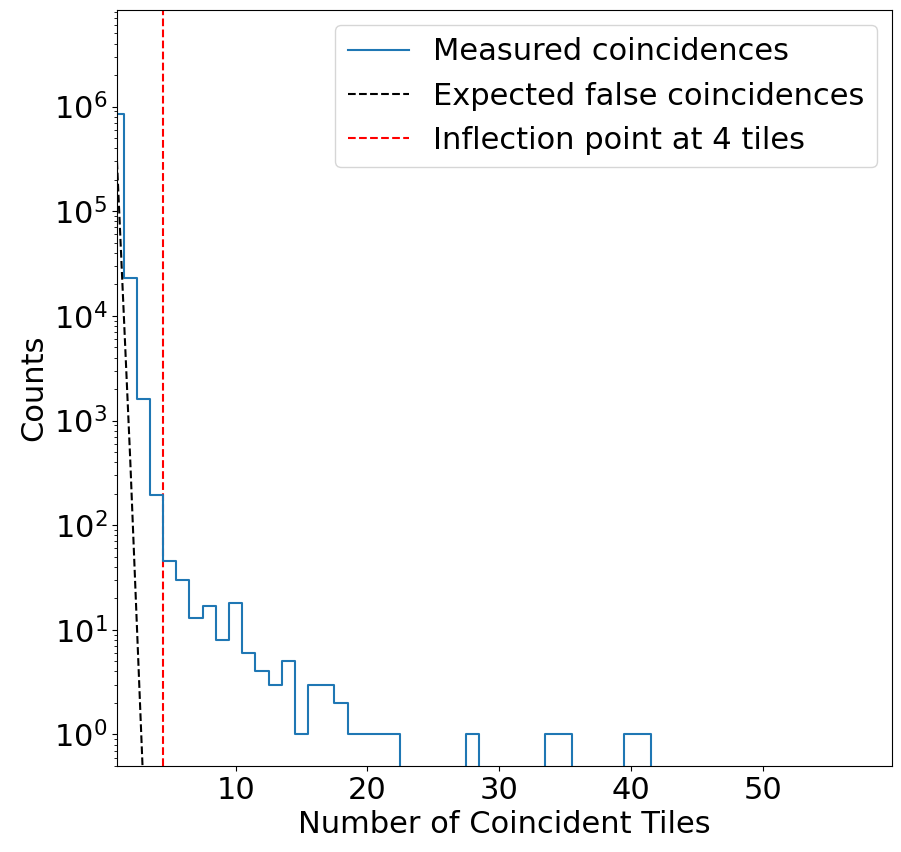}
    \caption{A histogram of the number of coincident tiles (window of 17.7\,$\mathrm{\mu s}$) per detected signal for an hour observation. The dashed line is the expected rate of coincidence from thermal noise.}
    \label{fig:coinc}
\end{figure}

To study these RFI events in more detail, we first identified all $6\,\sigma$ events, i.e.\ sample excursions with magnitude greater than six times the root-mean-square (RMS)  sample value in each tile/polarisation. We then grouped all samples within a coincidence window of 17\,$\mu$s, corresponding approximately to the light travel time across the MWA. Each event was characterised primarily by the number of tiles, $N_{\rm coinc}$, with excursions within the coincidence window.

Figure \ref{fig:coinc} shows the distribution of $N_{\rm coinc}$
as well as the predicted distribution from Gaussian noise at a $6\,\sigma$ threshold. As expected from Figure~\ref{fig:gaussianness}, the excess of high sample values leads to $N_{\rm coinc}$ significantly exceeding predictions. There is an inflection point near $N_{\rm coinc}=$4--5, indicating that detections with five or more coincident tiles are likely to be dominated by signals that cannot be attributed to thermal noise. These signals will either be RFI or radio emission from cosmic ray air showers. Following an extensive search, the majority of the signals that contribute to the excess of $6\,\sigma$ detections in the $N_{\rm coinc}=$1--3 range appear to be random and uncorrelated, rather than due to locally generated RFI.

The total number of RFI events will be less than the total number of events with $N_{\rm coinc}\ge4$ --- 363 events in this sample, i.e.\ one event every 10 seconds. Due to the prevalence of RFI, most other experiments that search for cosmic ray events use a particle detector array to act as a trigger \citep{LOFAR_detecting_cosmic_rays,AERA_Energy_Estimation}. Those without such a trigger typically suffer from very high rates of coincident RFI. For example, \citet{OVRO-LWA} used radio data only to identify 10 cosmic ray events in a 40\,hr window. However, the raw rate of candidate events was $500$\,Hz, with the resulting RFI rejection cuts leading to a loss of 55\% of the total observation time. In a related experiment to use the Parkes radio telescope to look for nanosecond radio pulses produced by ultra-high-energy particles hitting the Moon, \citet{LUNASKA_Parkes} recorded a raw rate of 250\,Hz RFI events, which was reduced to 1.6\,Hz using an anti-coincidence filter. That the MWA sees a much lower rate of RFI events will make the identification of cosmic ray events significantly easier, and is testament to the low-RFI environment of the MRO.

The cosmic ray event rate over the MWA core should be similar to that of LOFAR and OVRO-LWA, with a rate of approximately one per hour \citep{LOFAR_detecting_cosmic_rays,OVRO-LWA}. A unique characteristic of cosmic ray signals is their localised ground pattern, with emission concentrated in a ring of a few hundred metres diameter, which gets projected onto the ground plane according to the cosmic ray arrival direction \citep{HuegeReview}. Far-field RFI, on the other hand, should be detected across the array, and be readily distinguishable from cosmic ray events. Any locally generated RFI may appear as a localised transient signal originating from the ground, and should be distinguishable using our proven ability to localise RFI sources. The expected East-West polarisation of cosmic ray events is a further discriminant that can be used to identify these signals. We therefore plan to use these techniques to search for cosmic ray events in a much larger data sample in the near future.

\section{Conclusion}

We have successfully implemented a method to reconstruct Murchison Widefield Array data at inverse bandwidth (16.2\,ns) resolution. The necessity of inverting two stages of polyphase filterbanks results in the average loss of 9.5\% in the peak of an impulse. The method has been tested using a calibrator pulse, allowing the location of the calibrator to be reconstructed at the expected resolution of 5\,m. This ultra-high time-resolution mode is necessary and sufficient for detecting cosmic rays, which produce impulse-like radio signals in the MWA frequency range. A preliminary analysis of the nanosecond-scale RFI environment using one hour's worth of data at the MRO indicates a very low rate of impulsive RFI, of order 0.1\,Hz, which should readily allow the characteristic ground pattern of cosmic ray events to be identified. This motivates a search for cosmic ray events with a much longer observation period, and also suggests that this experiment will be feasible with the Square Kilometre Array, the low-frequency component of which will also be deployed at the Murchison Radio-astronomy Observatory.

\section*{Acknowledgments}

The authors would like to thank Nichole Barry, Aman Chokshi, David Emrich, Jack Line, Andrew McPhail,  Adrian Sutinjo, and Andrew Williams for their assistance in performing the gas lighter experiment. This scientific work makes use of the Murchison Radio-astronomy Observatory, operated by CSIRO. We acknowledge the Wajarri Yamatji people as the traditional owners of the Observatory site. Support for the operation of the MWA is provided by the Australian Government (NCRIS), under a contract to Curtin University administered by Astronomy Australia Limited. We acknowledge the Pawsey Supercomputing Centre which is supported by the Western Australian and Australian Governments. This research was supported by the Australian Government through the Australian Research Council's Discovery Projects funding scheme (project DP200102643). A.~Williamson would like to thank Innovation Central Perth, a collaboration of Cisco, Curtin University, Woodside and CSIRO's Data61, for their PhD scholarship. This research made use of Python libraries \textsc{Matplotlib} \citep{Matplotlib2007}, \textsc{NumPy} \citep{Numpy2020}, \textsc{SciPy} \citep{SciPy2019}, and \textsc{Astropy} \citep{astropy:2013,astropy:2018}, and NASA's Astrophysics Data System Bibliographic Services.

\bibliographystyle{ws-jai}
\bibliography{bibliography}

\end{document}